\documentclass[11pt]{article}

\usepackage{hyperref}
\usepackage{amsmath,amsfonts,amssymb}
\hypersetup{dvips,dvipdfm,colorlinks=true,urlcolor=magenta,filecolor=magenta,linktoc=page,citecolor=red,linkcolor=blue,bookmarks=true}
\usepackage{graphicx,epstopdf}
\usepackage{multirow,array}
\begin{document}
	\title{\textbf{A study on existence of Wormhole in Lemaitre-Tolman-Bondi model}}
	\author{Subenoy Chakraborty \footnote{\url{schakraborty.math@gmail.com} (corresponding author)}~~and~~
		Madhukrishna Chakraborty\footnote{\url{chakmadhu1997@gmail.com}}
		\\ \textit{Department of Mathematics, Jadavpur University, Kolkata - 700032, India}}
	\date{}
	\maketitle
	\begin{abstract}
		An investigation has been done for possible existence of evolving wormhole (WH) solution in the background of inhomogeneous Lemaitre-Tolman-Bondi (LTB) space-time geometry. Using separable product form of the geometric (or area) radius, the shape function can be evaluated in terms of the functional part of the radial coordinate in the area radius. The conditions for a viable WH configuration have been examined.
		\end{abstract}
		\maketitle
	\section{Introduction}
	 A popular hypothetical astrophysical object, widely used in the literature and suitable for scientific friction is the Wormhole (WH) geometry. It is considered as a hypothetical bridge connecting two distant universes or two asymptotic regions of the same universe. The notion of such interconnecting bridges between space-times was introduced by Einstein and Rosen (known as Einstein-Rosen bridge) \cite{Einstein:1935tc}. However, the term ``wormhole" was first put forward in literature by Misner and Wheeler \cite{Misner:1957mt}. More importantly, the pioneering work about traversability of the WH was done by Morris and Thorne \cite{Morris:1988cz} for static spherically symmetric space-time in Einstein gravity. They have shown that regular red-shift function (i.e, without any horizon) is necessary for the traversability of the WH. On the other hand, the matter field used for the WH geometry must violate the null energy condition (NEC) in the vicinity of the throat \cite{Ellis:1973yv}-\cite{Ida:1999an}. Consequently, other energy conditions are also violated near the throat. As classical or usual matter satisfies all the energy conditions so existence of traversable WH demands exotic matter at least in the vicinity of the throat. So far, majority of works on WH physics are related to static space-time background mostly in Einstein gravity and in modified theories of gravity. The investigation of the dynamical (or evolving) WH \cite{Cataldo:2008pm}-\cite{Cataldo:2008ku} has been initiated in recent past and they are mostly confined to inhomogeneous FLRW space-time background. The present work is a further extension where evolving WH geometry has been considered in the background of inhomogeneous LTB model \cite{Lemaitre:1933gd}, \cite{Bondi:1947fta}. 
	 
	 In General Relativity (GR), one of the very well known spherically symmetric model is the LTB space-time geometry which was formulated by Lemaitre, Tolman and Bondi  \cite{Bondi:1947fta}. It is one of the simplest, inhomogeneous and anisotropic cosmological model which is used not only for studying some new ideas in cosmology but also to interpret the recent observational data \cite{Celerier:2006hu}-\cite{Clarkson:2010uz}. It is speculated that presence of local inhomogeneity around us may have effect on the interpretation of cosmological data. Most of the efforts for estimating these effects so far deal with some ansatz for the profile of the inhomogeneity \cite{Clarkson:2010uz} and then the effects on cosmological observations has been calculated numerically. The limitation of such an approach depends on the particular functional form for modelling the local inhomogeneity as well as complete dependence on numerical evaluation. To resolve these issues, a general study has been done analytically and it is possible to derive luminosity \cite{Romano:2012gk}, \cite{Grande:2011hm} distance relation at low red-shift for an observer at the center of LTB model. The structure of the paper is as follows: The basic features of LTB model has been presented in Section 2. The formation of WH in the background LTB model has been given in Section 3. Also, viability of the WH solution and cosmic evolution of the background geometry has been studied in this section. A brief discussion and conclusion about the WH geometry is presented at the end of Section 4.
	 \section{Basic features of LTB space-time model}
	 A spherically symmetric space-time is considered having radial inhomogeneities as observed from the center. The spatial co-ordinates are assumed to be co-moving (i.e, $\dfrac{dx^{i}}{dt}=0$) with the matter having spatial origin ($x^{i}=0$) as the center of symmetry. Also, the time co-ordinate $(x^{0}=t)$ measures the proper time of the co-moving fluid. Now, the general form of the line element for such spherically symmetric model can be chosen as \cite{Moffat:2005ii}
	 \begin{equation}
	 	ds^{2}=-dt^{2}+X^{2}(r,t)dr^{2}+Y^{2}(r,t)d\Omega_{2}^{2}\label{eq1}
	 \end{equation} with $d\Omega_{2}^{2}=d\theta^{2}+\sin^{2}\theta d\phi^{2}$, the metric on unit 2-sphere. It should be noted that the line element (\ref{eq1}) reduces to the usual FLRW line element as a special case by choosing $X$ and $Y$ in separable form as $X(r,t)\rightarrow \dfrac{a(t)}{1-\kappa r^{2}}$ and $Y(r,t)\rightarrow a(t)r$. The matter component is assumed to have two fluids:
 Fluid 1. Inhomogeneous and anisotropic in nature: $(\rho_{1}(r,t),~p_{r}(r,t),~p_{t}(r,t))$ and
 Fluid 2. Homogeneous and isotropic in nature: $(\rho_{2}(t),~p_{2}(t),~p_{2}(t))$. So, the explicit form of Einstein's field equations $G_{\mu\nu}=8\pi G T_{\mu\nu}$ can be written as
 \begin{eqnarray}
 	-2\dfrac{Y''}{YX^{2}}+2\dfrac{Y'}{Y}\dfrac{X'}{X^{3}}+2\dfrac{\dot{X}}{X}\dfrac{\dot{Y}}{Y}+\dfrac{1}{Y^{2}}+\left(\dfrac{\dot{Y}}{Y}\right)^{2}-\left(\dfrac{Y'}{XY}\right)^{2}=8\pi G (\rho_{1}+\rho_{2})\label{eq2}\\
 	2\dfrac{Y''}{Y}+\dfrac{\dot{Y}^{2}}{Y^{2}}+\dfrac{1}{Y^{2}}-\left(\dfrac{Y'}{XY}\right)^{2}=-8\pi G(p_{r}+p_{2})\label{eq3}\\
 	-\dfrac{Y''}{X^{2}Y}+\dfrac{\ddot{Y}}{Y}+\dfrac{\dot{X}}{X}\dfrac{\dot{Y}}{Y}+\dfrac{X'}{X^{3}}\dfrac{Y'}{Y}+\dfrac{\ddot{X}}{X}=-8\pi G(p_{t}+p_{2})\label{eq4}
 \end{eqnarray} and 
\begin{equation}
	\dfrac{\dot{Y}'}{Y'}=\dfrac{\dot{X}}{X}\label{eq5}
\end{equation}
Here, the notations over dot and over dash represent derivatives with respect to time and radial coordinate respectively. It is to be noted that,  out of the above four field equations namely (\ref{eq2})-(\ref{eq5}), only three are independent and the rest one can be obtained from the other three. The first integral of (\ref{eq5}) gives $X$ as 
\begin{equation}
	X=Y'(r,t).C(r)
\end{equation} with $C(r)$ an arbitrary integration function. Now, choosing $C(r)=\dfrac{1}{\sqrt{1-\kappa(r)}}$ with $\kappa(r)<1$, the usual form of the LTB line element can be written as (replacing $Y$ by $R$ for convenience)
\begin{equation}
	ds^{2}=-dt^{2}+\dfrac{R'^{2}}{1-\kappa(r)}dr^{2}+R^{2}d\Omega_{2}^{2}\label{eq7}
\end{equation}
It is to be noted that the arbitrary function $\kappa(r)$ is related to the curvature of the $t=$ constant hyper-surface and $R$ is termed as area radius of the spherical surface. In fact, $\kappa(r)$ classifies the space-time as bounded, marginally bounded or unbounded according as $\kappa(r)>0$, $\kappa(r)=0$ or $\kappa(r)<0$. Using these modifications, the explicit field equations for the LTB line element (\ref{eq7}) are given by
\begin{eqnarray}
	2\dfrac{\dot{R}'}{R'}\dfrac{\dot{R}}{R}+\dfrac{\dot{R}^{2}}{R^{2}}+\dfrac{\kappa'}{RR'}+\dfrac{\kappa}{R^{2}}=\rho_{1}(r,t)+\rho_{2}(t)\label{eq8}\\
	2\dfrac{\ddot{R}}{R}+\dfrac{\dot{R}^{2}}{R^{2}}+\dfrac{\kappa}{R^{2}}=-\left(p_{r}(r,t)+p_{2}(t)\right)\label{eq9}\\
	\dfrac{\ddot{R}'}{R'}+\dfrac{\ddot{R}}{R}+\dfrac{\dot{R}'}{R'}\dfrac{\dot{R}}{R}+\dfrac{r\kappa'}{2RR'}=-\left(p_{t}(r,t)+p_{2}(t)\right)\label{eq10}
\end{eqnarray}
Now, the conservation of the energy-momentum tensor for both the fluids take the following explicit form as
\begin{eqnarray}
	\dot{\rho_{1}}(r,t)+(\rho_{1}+p_{r})\dfrac{\dot{R}'}{R'}+2(\rho_{1}+p_{t})\dfrac{\dot{R}}{R}=0\label{eq11}\\
	p_{r}'+2\dfrac{R'}{R}(p_{r}-p_{t})=0\label{eq12}\\
	\dot{\rho_{2}}+\left(\dfrac{\dot{R}'}{R'}+2\dfrac{\dot{R}}{R}\right)(\rho_{2}+p_{2})=0\label{eq13}
\end{eqnarray} with
\begin{equation}
	3H=\left(\dfrac{\dot{R}'}{R'}+2\dfrac{\dot{R}}{R}\right),
\end{equation} the average Hubble parameter. Further, the space-time geometry can be considered as successive shells labeled by $r$ and $R(r,t)$ stands for location of the shell marked by $r$ at time $t$ having an initial condition (by proper rescaling)
\begin{equation}
	R(r,0)=r
	\end{equation}
Finally, in the LTB model the proper volume of a sphere of radius $R$ at $t=$ constant hyper-surface can be written in co-moving coordinates in the form \cite{Chakraborty:2013tra}
\begin{equation}
	V_{p}=\int_{0}^{2\pi}d\phi\int_{0}^{\pi}d\theta\int_{0}^{r}dr\sqrt{^{(3)}g}
\end{equation} where $^{(3)}g$ is the determinant of the $t=$constant, $3D$ hyper-surface having expression
\begin{equation}
	^{(3)}g=\dfrac{(R')^{2}R^{4}}{1-\kappa(r)}\sin^{2}\theta
\end{equation}
\section{Formulation of WH configuration in LTB model}
To construct WH configuration in LTB model one has to restrict to bounded LTB model with $\kappa(r)=\dfrac{b(r)}{r}$, so that the LTB model becomes 
\begin{equation}
	ds^{2}=-dt^{2}+\dfrac{R'^{2}}{1-\dfrac{b(r)}{r}}dr^{2}+R^{2}(d\theta^{2}+\sin^{2}\theta d\phi^{2})\label{eq18}
\end{equation}
To have a traversable WH geometry, Morris- Throne \cite{Morris:1988cz} introduced some restrictions on the shape function $b(r)$ namely,
\begin{enumerate}
	\item The function $b(r)$ should be such that it may connect two asymptotically flat regions with a minimum value of $b(r)$ at the junction, the throat. As $b(r)$ is indicative of the shape at the WH throat so it is termed as shape function. Mathematically, at the throat ($r=r_{0}$), $b(r)$ should be restricted as $b(r_{0})=r_{0}$ and $b'(r_{0})<1$ \cite{Ghosh:2021msy}. 
	\item To maintain the space-like behavior of $r$, one must have
	$b(r)<r$ for $r>r_{0}$.
	\item Asymptotic flatness: $\dfrac{b(r)}{r}\rightarrow0$ as $r\rightarrow+\infty$ \cite{Dutta:2024luw}.
	\item The constraint of minimum radius at the throat with the traversability criterion impose huge tension at the throat. This is balanced by the violation of null energy condition (NEC) at the throat i.e, $\rho+p_{r}<0$ i.e, $\dfrac{rb'-b}{r^{3}}<0$, an essential feature for traversability of the WH \cite{Halder:2019urh}.
	\end{enumerate}
Thus, the field equations (\ref{eq8})-(\ref{eq10}) for the WH configuration (\ref{eq18}) modify to
\begin{eqnarray}
		2\dfrac{\dot{R}'}{R'}\dfrac{\dot{R}}{R}+\dfrac{\dot{R}^{2}}{R^{2}}+\dfrac{(rb'-b)}{r^{2}RR'}+\dfrac{b(r)}{rR^{2}}=\rho_{1}(r,t)+\rho_{2}(t)\label{eq19}\\
		2\dfrac{\ddot{R}}{R}+\dfrac{\dot{R}^{2}}{R^{2}}+\dfrac{b(r)}{rR^{2}}=-p_{r}(r,t)-p_{2}(t)\label{eq20}\\
			\dfrac{\ddot{R}'}{R'}+\dfrac{\ddot{R}}{R}+\dfrac{\dot{R}'}{R'}\dfrac{\dot{R}}{R}+\dfrac{(rb'-b)}{2r^{2}RR'}=-p_{t}(r,t)-p_{2}(t)\label{eq21}
\end{eqnarray}
Now using $R(r,t)=A(r)B(t)$ for convenience we obtain from equation (\ref{eq19}),
\begin{eqnarray}
	\dfrac{1}{B^{2}(t)}\left(\dfrac{(rb'-b)}{r^{2}AA'}+\dfrac{b}{rA^{2}}\right)=\rho_{1}(r,t)\label{eq22}\\
	3\dfrac{\dot{B}^{2}}{B^{2}}=\rho_{2}(t)\label{eq23}
\end{eqnarray}
From equation (\ref{eq20}), we get
\begin{equation}
	\dfrac{b}{r A^{2}B^{2}}=-p_{1r}(t,r)\label{eq24}
\end{equation} and 
\begin{equation}
	2\dfrac{\ddot{B}}{B}+\dfrac{\dot{B}^{2}}{B^{2}}=-p_{2}(t)\label{eq25}
\end{equation}
Also, from equation (\ref{eq21}) we obtain in separable form as
\begin{equation}
	\dfrac{(rb'-b)}{2r^{2}AA'B^{2}}=-p_{1t}(r,t)\label{eq26}
\end{equation} and same as equation (\ref{eq25}). Now subtracting equation (\ref{eq26}) and (\ref{eq24}), the difference between the radial and transverse pressure components has the expression
\begin{equation}
	p_{1t}-p_{1r}=-\dfrac{A}{2A'B^{2}}\dfrac{\partial}{\partial r}\left(\dfrac{b}{A^{2}r}\right)\label{eq27}
\end{equation}
Using (\ref{eq27}) in the conservation equation (i.e, the relativistic Euler equation) (\ref{eq12}) and after integrating once one may get the radial pressure component as 
\begin{equation}
	p_{1r}(r,t)=-\dfrac{b(r)}{rA^{2}B^{2}}+p_{0}(t)\label{eq28}
\end{equation}. Putting (\ref{eq28}) in (\ref{eq27}) one has 
\begin{equation}
	p_{1t}(r,t)=\dfrac{(b-rb')}{2r^{2}AA'B^{2}}+p_{0}(t)\label{eq28*}
\end{equation} Here $p_{0}(t)$ is an arbitrary integration function. It is to be noted that in order to maintain or preserve  the anisotropic nature of Fluid I, the function $A(r)$ should not be of the form $A^{2}(r)=a_{0}\left(\dfrac{b(r)}{r}\right)$ for some constant $a_{0}$. Further, if the anisotropic pressure components are assumed to be linearly related i.e,
\begin{equation}
	p_{1t}=\alpha p_{1r}~~~(\alpha \neq 1)\label{eq29}
\end{equation} then the shape function $b(r)$ takes the form
\begin{equation}
	\dfrac{b(r)}{r}=b_{0}A^{2\alpha}+\dfrac{p_{0}}{(\alpha-1)}A^{2}\label{eq31}
\end{equation} where $b_{0}$ is a constant of integration and the constant $p_{0}$ is related to the arbitrary function $p_{0}(t)$ by the relation
\begin{equation}
	p_{0}(t)=\dfrac{p_{0}}{(\alpha-1)B^{2}}
\end{equation} where $p_{0}$ is a constant. Now substituting equations (\ref{eq28}) and (\ref{eq28*}) in the conservation equation (\ref{eq11}), the energy density for the anisotropic fluid takes the form
\begin{equation}
	\rho_{1}(r,t)=\dfrac{b_{0}(2\alpha+1)}{B^{2}}A^{2\alpha-2}+\dfrac{\rho_{0}(r)}{B^{3}}\label{eq33}
\end{equation} with $\rho_{0}(r)$ an arbitrary integration function. If the explicit form of the shape function from (\ref{eq31}) is used in the equations (\ref{eq28}) and (\ref{eq28*}), then the simplified form of the anisotropic fluid's pressure components are
\begin{equation}
	p_{1r}=-\dfrac{b_{0}A^{2(\alpha-1)}}{B^{2}},~p_{1t}=-\dfrac{\alpha b_{0}A^{2(\alpha-1)}}{B^{2}}\label{eq34}
\end{equation}
If the above explicit form for energy density (given by (\ref{eq33})) and pressure components given by (\ref{eq34}) for the anisotropic fluid (i.e, fluid I) are substituted into the field equations (\ref{eq19}) and (\ref{eq20}), the energy density and isotropic pressure for Fluid II are given by
\begin{eqnarray}
	\rho_{2}(t)=3\dfrac{\dot{B}^{2}}{B^{2}}+\dfrac{3p_{0}(\alpha-1)}{B^{2}}+\dfrac{\rho_{0}(r)}{B^{3}}\label{eq35}\\
	p_{2}(t)=-2\dfrac{\ddot{B}}{B}-\dfrac{\dot{B}^{2}}{B^{2}}-\dfrac{p_{0}(\alpha-1)}{B^{2}}\label{eq36}
\end{eqnarray} For consistency of equation (\ref{eq35}), $\rho_{0}(r)=\rho_{0}$, a constant while the conservation equation (\ref{eq13}) implies $\rho_{0}=0$.  Moreover, for the present LTB model the line element takes the form
\begin{equation}
	ds^{2}=-dt^{2}+B^{2}(t)\left(\dfrac{dX^{2}}{g(X)}+X^{2}d\Omega_{2}^{2}\right)\label{eq37**}
\end{equation}
where $X=A(r)$ and $g(X)=1-\dfrac{b(r)}{r}=1-b_{0}X^{2\alpha}-\dfrac{p_{0}}{(\alpha-1)}X^{2}$. This metric is analogous to the FLRW metric with $g(X)$ related to the curvature of the space-time.

\begin{figure}[h!]
	\begin{minipage}{0.3\textwidth}
		\centering\includegraphics[height=5cm,width=6cm]{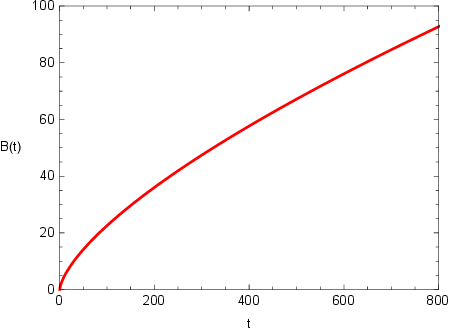}
	\end{minipage}~~~~~~~~~~~~~~~~~~~~~~~~~~~~~~~~~~~~~~~~~~~~~~~~~~~
	\begin{minipage}{0.3\textwidth}
		\centering\includegraphics[height=5cm,width=6cm]{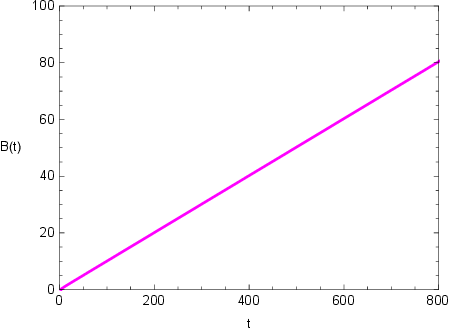}
	\end{minipage}
	\begin{minipage}{0.3\textwidth}
		\centering\includegraphics[height=5cm,width=6cm]{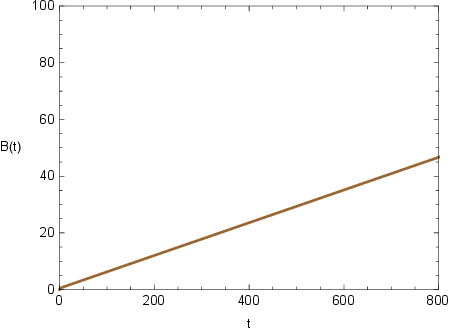}
	\end{minipage}~~~~~~~~~~~~~~~~~~~~~~~~~~~~~~~~~~~~~~~~~~~~~~~~~~~~~~
	\begin{minipage}{0.3\textwidth}
		\centering\includegraphics[height=5cm,width=6cm]{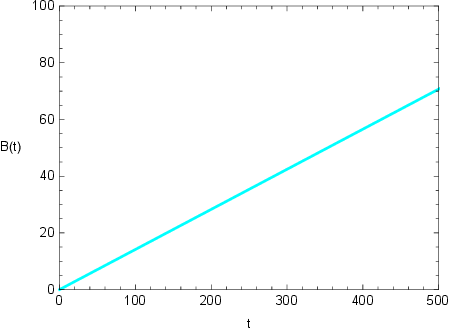}
	\end{minipage}
\centering
\begin{minipage}{0.3\textwidth}
	\centering\includegraphics[height=5cm,width=6cm]{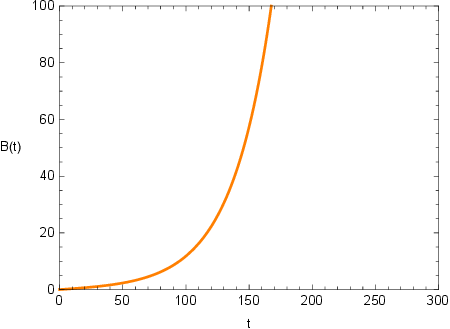}
\end{minipage}
	\caption{\small[$B(t)$ vs $t$ for various choices of the parameters namely: (i)Red curve: $\omega_{0}=0,~p_{0}=0.001,~\alpha=0.001,~B_{0}=0.5,~t_{0}=0.5$; (ii) Magenta curve: $\omega_{0}=\dfrac{1}{3},~p_{0}=0.01,~\alpha=0.01,~B_{0}=0.001,~t_{0}=0.001$; (iii) Brown curve: $\omega_{0}=1,~p_{0}=0.01,~\alpha=0.001,~B_{0}=0.001,~t_{0}=0.001$; (iv) Cyan curve: $\omega_{0}=-\dfrac{1}{3},~p_{0}=0.01,~\alpha=0.001,~B_{0}=0.01,~t_{0}=0.001$; (v) Orange curve: $\omega_{0}=-1,~p_{0}=0.001,~\alpha=0.001,~B_{0}=0.001,~t_{0}=0.0001$]}\label{f1}
\end{figure}

 Now, to determine the background cosmic evolution we assume the isotropic pressure to be barotropic in nature i.e, $p_{2}(t)=\omega \rho_{2}(t)$ with $\omega$, the barotropic index. So, the evolution equation becomes 
\begin{equation}
	2\dfrac{\ddot{B}}{B}+(1+3\omega)\dfrac{\dot{B}^{2}}{B^{2}}+\dfrac{(1+3\omega)p_{0}}{(\alpha-1)B^{2}}=0\label{eq37}
\end{equation}
We now consider the following cases:\\
\textbf{Case-1} $\omega=\omega_{0},$ a constant, $p_{0}=0$. In this case, $B$ evolves in  power law fashion for $\omega_{0}\neq -1$ and in exponential form for $\omega_{0}=-1$  given as
\begin{equation}
	B(t)=
	\begin{cases}
		B_{0}t^{\mu},~\mu=\dfrac{2}{3(1+\omega_{0})} & \text{for } \omega_{0}\neq-1\\
		B_{0}\exp(\alpha_{0} t)& \text{for } \omega=-1
	\end{cases}\label{eq26**}
\end{equation}  where $B_{0},~\alpha_{0}$ are arbitrary integration constants.\\
\textbf{Case-2} $\omega$ is variable, $p_{0}=0$\\
If modified chaplygin gas is chosen as the isotropic fluid with equation of state \cite{Debnath:2004cd}
\begin{equation}
	p_{2}=-\rho_{2}-\dfrac{\mu}{\rho_{2}^{n}},
\end{equation} $\mu$, $n$ are constants then from the conservation equation 
\begin{equation}
	\dot{\rho_{2}}+3\dfrac{\dot{B}}{B}(\rho_{2}+p_{2})=0\label{eq40}
\end{equation} one gets after integration
\begin{equation}
	\rho_{2}=\left(\rho_{c}+3\mu_{0}\ln B\right)^{\frac{1}{(n+1)}}
\end{equation} Here $\rho_{c}$ is a constant of integration and $\mu_{0}=\mu(n+1)$. Thus, the equation of state parameter now becomes 
\begin{equation}
	\omega=\dfrac{p_{2}}{\rho_{2}}=-1-\dfrac{\mu}{\rho_{2}^{n+1}}=-1-\dfrac{\mu}{(\rho_{c}+3\mu_{0}\ln B)}
\end{equation}
Using this $\omega$, one can solve the evolution equation to have
\begin{equation}
	B=B_{c}\exp(\delta(t-t_{0})^{n})
\end{equation} with $B_{c}$, $t_{0}$ and $\delta$ are integration constants.\\
\textbf{Case-3} $\omega=\omega_{0}$, $p_{0}\neq0$\\
For this case, the evolution equation (\ref{eq40}) has a first integral
\begin{equation}
	\dot{B}^{2}=\dfrac{p_{0}}{1-\alpha}+B_{0}B^{-(1+3\omega_{0})}
\end{equation} with $B_{0}$, a constant of integration. Thus the solution for $B$ can be written as an integral form as 
\begin{equation}
	\int \dfrac{B^{\frac{1+3\omega_{0}}{2}}dB}{\sqrt{B_{0}-\left(\frac{p_{0}}{\alpha-1}\right)B^{1+3\omega_{0}}}}=(t-t_{0})
\end{equation}
The explicit solutions for different choices of $\omega_{0}$ can be written as
\begin{equation}
	\begin{cases}
		(t-t_{0})=\dfrac{(1-\alpha)}{p_{0}}\sqrt{B\left(B_{0}-\left(\dfrac{p_{0}}{\alpha-1}\right)B\right)}+B_{0}\left(\dfrac{(\alpha-1)}{p_{0}}\right)^{\frac{3}{2}}\sin^{-1}\left(\sqrt{\dfrac{p_{0}B}{(\alpha-1)B_{0}}}\right) & \text{for } \omega_{0}=0\\
		(t-t_{0})=\dfrac{1-\alpha}{p_{0}}\sqrt{B_{0}-\dfrac{p_{0}}{\alpha-1}B^{2}}~or ~B(t)=\left(\left(\dfrac{\alpha-1}{p_{0}}\right)\left(B_{0}-\left(\dfrac{p_{0}}{1-\alpha}\right)(t-t_{0})^{2}\right)\right)^{\frac{1}{2}}& \text{for } \omega_{0}=\dfrac{1}{3}\\
		(t-t_{0})=\dfrac{1}{\sqrt{3B_{0}}}B^{3}2^{F_{1}}\left(\dfrac{1}{2},~\dfrac{3}{4},~\dfrac{7}{4},~\dfrac{p_{0}B^{4}}{(\alpha-1)B_{0}}\right)& \text{for } \omega_{0}=1\\
		(t-t_{0})=\left(B_{0}-\dfrac{p_{0}}{\alpha-1}\right)^{-\frac{1}{2}}B(t)& \text{for } \omega_{0}=-\dfrac{1}{3}\\
		(t-t_{0})=\dfrac{1}{\sqrt{B_{0}}}\tanh^{-1}\left(\dfrac{\sqrt{B_{0}}B}{\sqrt{B_{0}B^{2}-\left(\dfrac{p_{0}}{\alpha-1}\right)}}\right)& \text{for } \omega_{0}=-1
	\end{cases}
\end{equation} 
The variation of $B(t)$ for different choices of $\omega_{0}$ and the parameters ($t_{0},~p_{0},~B_{0},~\alpha$) are shown graphically in FIG. (\ref{f1}).
 The analogy of LTB model with FLRW has been done previously in equation (\ref{eq37**}). Thus, we find that $B(t)$ in LTB model behaves like the cosmic scale factor $a(t)$ in FLRW case. The solution $B(t)$ for $\omega_{0}=0$ can not have an explicit form as a function of $t$, but the solution has singularity at $t=t_{0}$. The solutions $B(t)$ for $\omega_{0}=\pm \dfrac{1}{3}$ are in the power law form and there is a big-bang singularity at finite epoch for each case. The solution for $\omega_{0}=-1$ always corresponds to finite $B$ having a singularity at $t=t_{0}$ but from cosmological point of view it may correspond to the behavior of $a(t)$. For stiff fluid corresponding to $\omega_{0}=1$, $B$ is an increasing function of $t$ as shown in the FIG. \ref{f1} having big-bang singularity at $t=t_{0}$.

\begin{figure}[h!]
	\begin{minipage}{0.3\textwidth}
		\centering\includegraphics[height=6cm,width=7.5cm]{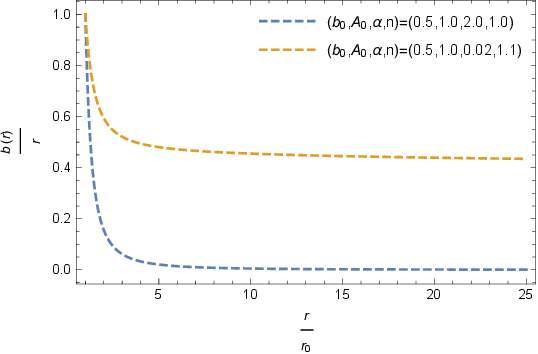}
	\end{minipage}~~~~~~~~~~~~~~~~~~~~~~~~~~~~~~~~
	\begin{minipage}{0.3\textwidth}
		\centering\includegraphics[height=6cm,width=7.5cm]{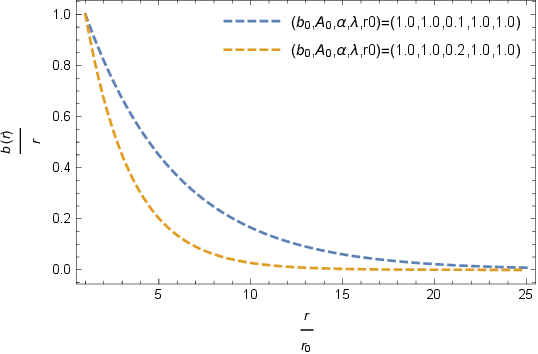}
	\end{minipage}
	\caption{\small $\dfrac{b(r)}{r}$ vs $\dfrac{r}{r_{0}}$ for various choices of the parameters involved for (i)$A(r)=A_{0}\left(\dfrac{r}{r_{0}}\right)^{-n}$ (left) and (ii) $A(r)=A_{0}\exp\left(-\lambda r_{0}\left(\dfrac{r}{r_{0}}-1\right)\right)$ (right).}\label{f2}
\end{figure}

\begin{figure}[h!]
		\begin{minipage}{0.3\textwidth}
		\centering\includegraphics[height=6cm,width=7.5cm]{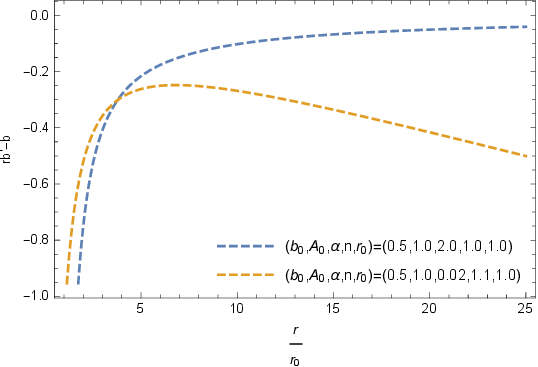}
	\end{minipage}~~~~~~~~~~~~~~~~~~~~~~~~~~~~~~~~
	\begin{minipage}{0.3\textwidth}
	\centering\includegraphics[height=6cm,width=7.5cm]{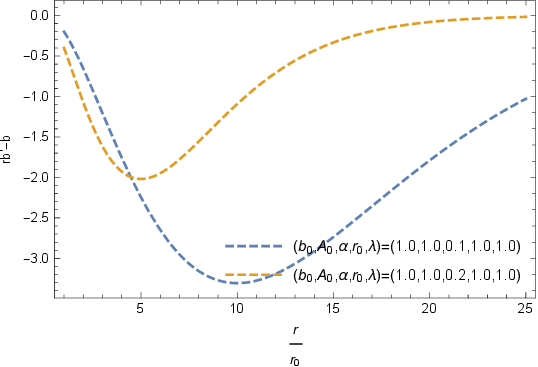}
\end{minipage}
\caption{\small $rb'-b$ vs $\dfrac{r}{r_{0}}$ for various choices of the parameters involved for (i)$A(r)=A_{0}\left(\dfrac{r}{r_{0}}\right)^{-n}$ (left) and (ii) $A(r)=A_{0}\exp\left(-\lambda r_{0}\left(\dfrac{r}{r_{0}}-1\right)\right)$ (right).}\label{f3}
	\end{figure}

We shall now discuss the feasibility of the WH solution  obtained in equation (\ref{eq31}). Using, the throat condition $b(r_{0})=r_{0}$ one may rewrite the shape function as
\begin{equation}
	\dfrac{b(r)}{r}=\left(\dfrac{A}{A_{0}}\right)^{2}+b_{0}A_{0}^{2\alpha}\left(\left(\dfrac{A}{A_{0}}\right)^{2\alpha}-\left(\dfrac{A}{A_{0}}\right)^{2}\right)\label{eq47}
\end{equation} with $A_{0}=A(r_{0})$.
The asymptotic flatness condition i.e, $\dfrac{b(r)}{r}\rightarrow0$ as $r\rightarrow0$ implies $A\rightarrow0$ as $r\rightarrow\infty$ ($\alpha>0$). Now, $\dfrac{b'r-b}{r^{2}}=\left(2\alpha b_{0}A^{2\alpha-1}+\dfrac{2p_{0}}{\alpha-1}\right)A'$. So, $b'r-b<0$ implies $A'<0$ i.e, $A$ is a decreasing function of $r$ for $r\geq r_{0}$ in order to satisfy the flare out condition. Therefore, we consider two choices of $A(r)$ namely, (i) $A(r)=A_{0}\left(\dfrac{r}{r_{0}}\right)^{-n}$; $n,A_{0}>0$ and (ii) $A(r)=A_{0}\exp(-\lambda (r-r_{0}))$; $\lambda>0$ so that the asymptotic flatness as well as flaring-out condition hold. For the first choice of $A(r)$, expression for $\dfrac{b(r)}{r}$ is given by
	\begin{equation}
	\dfrac{b(r)}{r}=\left(\dfrac{r}{r_{0}}\right)^{-2n}+b_{0}A_{0}^{2\alpha}\left(\left(\dfrac{r}{r_{0}}\right)^{-2\alpha n}-\left(\dfrac{r}{r_{0}}\right)^{-2n}\right)
	\end{equation}
and for the second choice of $A(r)$, the expression for $\dfrac{b(r)}{r}$ is given by
\begin{equation}
\dfrac{b(r)}{r}=\exp\left(-2\lambda r_{0} \left(\frac{r}{r_{0}}-1\right)\right)+b_{0}A_{0}^{2\alpha}\left(\exp\left(-2\alpha \lambda r_{0} \left(\frac{r}{r_{0}}-1\right)\right)-\exp\left(-2\lambda r_{0}\left(\frac{r}{r_{0}}-1\right)\right)\right)
\end{equation}
The expression for $(rb'-b)$ for both the choices of $A(r)$ are respectively given by
\scriptsize
\begin{eqnarray}
	rb'-b=-2 n r_{0}\left(\dfrac{r}{r_{0}}\right)^{1-2n}+2nr_{0}b_{0}A_{0}^{2\alpha}\left(\left(\dfrac{r}{r_{0}}\right)^{1-2n}-\alpha\left(\dfrac{r}{r_{0}}\right)^{1-2\alpha n}\right)\nonumber\\
	rb'-b=\left(\dfrac{r}{r_{0}}\right)^{2}r_{0}^{2}~\left(2\lambda(A_{0}^{2\alpha}b_{0}-1)\exp\left(-2\lambda r_{0}\left(\frac{r}{r_{0}}-1\right)\right)-2\alpha\lambda b_{0}A_{0}^{2\alpha}\exp \left(-2r_{0}\left(\frac{r}{r_{0}}-1\right)\alpha\lambda\right)\right)\nonumber
\end{eqnarray}
\normalsize
	The plots illustrating behavior of $\dfrac{b(r)}{r}$ vs $\dfrac{r}{r_{0}}$ are shown in figure FIG.(\ref{f2}). From the analytical expression as well as from the plots it is clear that $\dfrac{b(r)}{r}=1$ at $\dfrac{r}{r_{0}}=1$. Further, one may note that $\dfrac{b(r)}{r}\rightarrow 0$ as $r$ grows larger ensuring asymptotic flatness condition. Moreover, for both the choices of $A(r)$ the wormhole is infinitely extended as  $A$ is a decreasing function of $r$ for $r\geq r_{0}$ in order to satisfy the flare out condition. Graphically, the flaring-out property has also been shown in FIG. (\ref{f3}) and it is seen that for both the choices of $A(r)$ we have $(rb'-b)<0$ i.e, flare out condition holds.

\newpage
 We now examine, whether the energy conditions are satisfied or not for the present WH configuration. We analyze both the fluid components separately. The explicit inequalities for the validity of the energy conditions are
\begin{itemize}
		\item Null Energy Condition (NEC): $\rho+p_{r}\geq0$, $\rho+p_{t}\geq0$
	\item Weak Energy Condition (WEC): $\rho\geq0$, $\rho+p_{r}\geq0$, $\rho+p_{t}\geq0$
	\item Strong Energy Condition (SEC): $\rho+p_{r}\geq0$, $\rho+p_{t}\geq0$, $\rho+p_{r}+2p_{t}\geq0$
	\item Dominant Energy Condition (DEC): $\rho-|p_{r}|\geq0$, $\rho-|p_{t}|\geq0$
\end{itemize}

\newpage
To study the energy conditions explicitly in the present model we consider the following components of energy density and pressure
\begin{eqnarray}
	\rho(r,t)=\rho_{1}(r,t)+\rho_{2}(t)=\dfrac{b_{0}(2\alpha+1)}{B^{2}}A^{2\alpha-2}+\dfrac{\rho_{0}(r)}{B^{3}}+3\dfrac{\dot{B}^{2}}{B^{2}}+\dfrac{3p_{0}(\alpha-1)}{B^{2}}+\dfrac{\rho_{0}(r)}{B^{3}}\nonumber\\
	p_{r}=-\dfrac{b_{0}A^{2(\alpha-1)}}{B^{2}}-\dfrac{2\ddot{B}}{B}-\dfrac{\dot{B}^{2}}{B^{2}}-\dfrac{p_{0}(\alpha-1)}{B^{2}}\nonumber\\
	p_{t}=-\dfrac{\alpha b_{0}A^{2(\alpha-1)}}{B^{2}}-2\dfrac{\ddot{B}}{B}-\dfrac{\dot{B}^{2}}{B^{2}}-\dfrac{p_{0}(\alpha-1)}{B^{2}}
\end{eqnarray}
For fluid I, we have the following
\begin{eqnarray}
	\rho_{1}=\dfrac{b_{0}(2\alpha+1)}{B^{2}}A^{2\alpha-2}+\dfrac{\rho_{0}(r)}{B^{3}}\\
	p_{1r}=-\dfrac{b_{0}A^{2(\alpha-1)}}{B^{2}}\\
	p_{1t}=-\dfrac{\alpha b_{0}A^{2(\alpha-1)}}{B^{2}}
\end{eqnarray}
 Thus, we have
\begin{eqnarray}
	\rho_{1}+p_{1r}=\dfrac{2\alpha b_{0}}{B^{2}}A^{2(\alpha-1)}\\
	\rho_{1}+p_{1t}=\dfrac{b_{0}(\alpha+1)A^{2(\alpha-1)}}{B^{2}}\\
	\rho_{1}+p_{1r}+2p_{1t}=0\\
	\rho_{1}-p_{1r}=\dfrac{2b_{0}(\alpha+1)}{B^{2}}A^{2(\alpha-1)}\\
	\rho_{1}-p_{1t}=\dfrac{b_{0}(3\alpha+1)}{B^{2}}A^{2(\alpha-1)}
\end{eqnarray}For fluid I, it is easily found that all the energy conditions are satisfied if the anisotropy parameter $\alpha$ is chosen to be positive i.e, $p_{1r}$ and $p_{1t}$ are of same sign. Further, it is found that fluid II has no effect on the WH geometry, it only influences the cosmic evolution of the background space-time geometry. Since we are interested to examine the energy conditions near the throat so we have not examined the energy conditions for fluid II. In this context, it is worthy to mention that cosmological WH solution in LTB model has been presented in \cite{Bochicchio:2010df} where the authors considered WH shell of zero thickness joining two LTB universes with no radial accretion and similar to our result the material on the shell satisfies the energy conditions.
\section{Brief Discussion and Conclusion}
The present work investigates the possible WH configuration in the background of inhomogeneous LTB model. The existence of WH solution is possible only when the area radius is considered as the product of two functions- a function of time alone and a function of radial coordinate only. The shape function can not be obtained explicitly as function of radial co-ordinate rather it can be expressed only in a linear combination of two distinct power of the radial function (in the area radius). The viability of the WH geometry restricts the radial function (termed as area shape function) to be a decreasing function of the radial co-ordinate and it vanishes in the asymptotic limit. The WH solution so obtained may be considered as a family of possible WH configuration for different choices of the areal shape function. The cosmic evolution for different choices of the equation of state parameters have been shown both analytically as well as graphically. The results obtained from the present study can be summarized as follows:
\begin{itemize}
	\item A class of WH solutions are obtained in the polynomial form of the areal shape function with some restrictions for the viability of the WH geometry.
	\item The WH solution is possible for two non-interacting fluid components: Fluid I (inhomogeneous and anisotropic) and Fluid II (homogeneous and isotropic). A single inhomogeneous and anisotropic fluid will be inconsistent for the formulation of WH scenario.
	\item The WH formation is completely characterized by Fluid I only while Fluid II determines the cosmic evolution of the background space-time geometry only.
	\item The anisotropy parameter has an important role in characterizing the WH geometry. The WH solution is only possible for linear relation between the radial and transverse pressure components for Fluid I.
	\item As all the energy conditions are satisfied for fluid I so the present dynamic or evolving WH geometry does not need any exotic matter surrounding the throat.
\end{itemize} 
It will be interesting to consider interaction between two fluid components for possible WH scenario for future work.
	\section*{Acknowledgment}
The authors thank the anonymous reviewer for valuable and insightful comments that improved the quality and visibility of the paper. M.C thanks University Grants Commission (UGC) for providing the Senior Research Fellowship (ID:211610035684/JOINT CSIR-UGC NET JUNE-2021). S.C thanks FIST program of DST, Department of Mathematics, JU (SR/FST/MS-II/2021/101(C)). The authors are thankful to Inter University Centre for Astronomy and Astrophysics (IUCAA), Pune, India for their warm hospitality and research facilities where this work was carried out under the ``Visiting Research Associates" program of S.C.

	\end{document}